\documentclass[prl,aps,showpacs,onecolumn,floatfix]{revtex4}
\usepackage{epsfig} \usepackage{graphics} \usepackage{bm}
\usepackage{amssymb}
\usepackage{graphicx}
\begin{document}

\preprint{Lebed-2019}

\title{Breakdown of the equivalence between gravitational mass and
energy due to quantum effects}

\author{Andrei G. Lebed}

\affiliation{Department of Physics, University of Arizona, 1118 E. 4th Street,\\
Tucson, Arizona 85721, USA and\\
L.D. Landau Institute for Theoretical Physics, RAS, 2 Kosygina Street,\\
Moscow 117334, Russia\\
lebed@physics.arizona.edu}

\begin{abstract}
We review our recent theoretical results about inequivalence
between passive and active gravitational masses and energy in
semiclassical variant of general relativity, where gravitational
field is not quantized but matter is quantized. To this end, we
consider the simplest quantum body with internal degrees of
freedom - a hydrogen atom. We concentrate our attention on the
following physical effects, related to electron mass. The first
one is inequivalence between passive gravitational mass and energy
at microscopic level. Indeed, quantum measurement of gravitational
mass can give result, which is different from the expected, $m
\neq m_e + \frac{E_1}{c^2}$, where electron is initially in its
ground state; $m_e$ is the bare electron mass. The second effect
is that the expectation values of both passive and active
gravitational masses of stationary quantum states are equivalent
to the expectation value of energy. The most spectacular effects
are inequivalence of passive and active gravitational masses and
energy at macroscopic level for ensemble of coherent
superpositions of stationary quantum states. We show that, for
such superpositions, the expectation values of passive and active
gravitational masses are not related to the expectation value of
energy by the famous Einstein's equation, $m \neq \frac{E}{c^2}$.
In this review, we also improve several drawbacks of the original
pioneering works.
\end{abstract}



\maketitle

\section{1. Introduction}
Equivalence principle (EP) between gravitational and inertial
masses in a combination with the local Lorentz invariance of
spacetime is known to be a keystone of the classical general
relativity [1,2]. In the current scientific literature, there
exists a wide discussion if it can survive in the possible quantum
theory of gravity (see, for example, the recent Refs.[3-5]). Since
the quantum gravitation theory has not been elaborated yet, the EP
is often studied in framework of the so-called semiclassical
approach to quantum gravity, where gravitational field is not
quantized, but the matter is quantized [3-5]. Note that the EP for
a composite body is not a trivial notion even in general
relativity in the absence of quantum effects. Indeed, as shown in
Refs.[6-8], external gravitational field is coupled not directly
with energy of a composite body but with the combination
$R+3K+2P$, where $R$, $K$, and $P$ are rest, kinetic, and
potential energies, respectively. As mentioned in Ref.[8], and
considered in detail in Ref.[9], the above mentioned combination
can be changed into expected total energy, if we choose the proper
local coordinates, where the interval has the Minkowski's form.
Therefore, in classical general relativity passive gravitational
mass is equivalent to inertial one for a composite body [8,9], as
expected. On the other hand, as shown in
 Ref.[7], active gravitational mass of a composite classical
body is equivalent to its energy only after averaging the
gravitational mass over time. Semiclassical analysis [5] of the
Einstein's field equation has demonstrated that the expectation
values of active gravitational mass and energy are equivalent only
for stationary quantum states of a composite quantum body.
Situation is different for ensemble of coherent quantum
superpositions of stationary quantum states, where the expectation
values of active gravitational mass can oscillate in time [5] even
for superpositions with the constant expectation values of energy.
The results of
 Ref.[5] are against the equivalence of active gravitational
mass and energy even at macroscopic level in quantum gravity,
which has to modify the EP. Note that quantum effects also change
the status of the EP for passive gravitational mass of a quantum
body with internal degrees of freedom. As discussed in [10],
quantum effects break the equivalence of passive gravitational
mass and energy at a microscopic level. Let us consider this
phenomenon in more details. Suppose that there is a hydrogen atom
in ground state, $E_1$, and we switch on gravitational field to
measure electron mass, which is expected to be equal to $m = m_e
+\frac{E_1}{c^2}$, where $m$ is electron passive gravitational
mass, $m_e$ is the bare electron mass. Contrary to the above
mentioned common expectation, it has been shown in Ref.[10] that
quantum measurements of the mass can give the following results $m
= m_e + \frac{E_n}{c^2}$, where $E_n$ is energy of electron $nS$
orbital in a hydrogen atom, although the corresponding
probabilities are very small. Note that influence of quantum
effects on the EP is even more dramatic and, as shown in Ref.[11],
the equivalence between passive gravitational mass and energy is
broken even at a macroscopic level. Indeed, the above mentioned
equivalence exists for the expectation values of the mass and
energy only for stationary quantum states. In accordance with
results of
 Ref.[11], the equivalence between the expectation values of
passive gravitational mass and energy is broken for ensemble of
coherent quantum superpositions of stationary quantum states in a
hydrogen atom. Of course, all these statements are not restricted
by the atom but are common properties of any quantum body with
internal degrees of freedom.

\section{2. Goal}

In Sec. 3, we discuss that there is no equivalence between passive
gravitational mass and energy of electron in a hydrogen atom at a
microscopic level, using the local Lorentz invariance, which
defines electron wave functions in a gravitational field (we call
it method-1). We start from quantum state with a definite electron
energy in the absence of external gravitational field, $E_1$, and
show that quantum measurement of the mass in the field can give
values different from expected one, $m \neq m_e+\frac{E_1}{c^2}$,
although the corresponding probabilities are small. In Sec. 4, we
discuss the same results, using corrections to the Schr\"{o}dinger
equation for electron in a hydrogen atom, which contain the
so-called virial term in gravitational field. We stress importance
of the virial term for the breakdown of the above mentioned
equivalence (and, thus, for the breakdown of the EP) at a
microscopic level (we call this method-2). In Sec. 5, we discuss
the breakdown of the equivalence between passive gravitational
mass and energy (and, as a result - breakdown of the EP) at
macroscopic level. We show that this equivalence survives for
macroscopic ensemble of stationary quantum electron states in a
hydrogen atom due to the so-called quantum virial theorem. On the
other hand, it is also shown that for coherent ensembles of
superpositions of stationary quantum states the above mentioned
equivalence is not survived due to the quantum virial term. During
our calculations in Secs. 2-5, we use property of the local Lorentz
invariance of a spacetime in general relativity as well as
consider passive gravitational mass to be a quantity proportional
to weight of a composite body whose center of mass is fixed in
gravitational field by some forces of non-gravitational origin.
Finally, in Sec. 6, we discuss the EP between active gravitational
mass and energy for electron in a hydrogen atom at macroscopic
level. Our results are similar to that for passive gravitational
mass. Indeed, we show that the equivalence (and, thus, the EP)
survives for macroscopic ensembles of stationary quantum states,
whereas for macroscopic coherent ensembles of superpositions of
quantum states the equivalence is broken. In Sec. 7, we come to
the conclusion that the EP has to be seriously reformulated in the
presence of quantum effects in general relativity.

\section{3. Inequivalence of passive gravitational mass and energy
at microscopic level (method-1)}
\subsection{3.1. Electron wave function  in a hydrogen atom with a
definite energy in the absence of gravitational field} Suppose
that, at $t<0$, there is no gravitational field and electron is in
a ground state of a hydrogen atom, characterizing by the following
wave function:
\begin{equation}
\Psi_1(r,t) = \exp \biggl( \frac{-i m_e c^2 t}{ \hbar} \biggl)
\exp \biggl(\frac{-i E_1 t}{ \hbar} \biggl) \Psi_1(r),
\end{equation}
which is solution of the corresponding Schr\"{o}dinger equation:
\begin{equation}
i \hbar \frac{\partial \Psi_1(r,t)}{\partial t} = \biggl[ m_e c^2
- \frac{\hbar^2}{2m_e} \biggl( \frac{\partial^2}{ \partial x^2} +
\frac{\partial^2}{ \partial y^2} + \frac{\partial^2}{ \partial
z^2} \biggl) - \frac{e^2}{r} \biggl] \Psi_1(r,t).
\end{equation}
[Here $E_1$ is electron ground state energy, $r$ is distance
between electron with coordinates $(x,y,z)$ and proton; $\hbar$ is
the Planck constant, $c$ is the velocity of light.]

\subsection{3.2 Electron wave functions in a hydrogen atom in the
presence of gravitational field} At $t=0$, we perform the
following Gedanken experiment. We switch on a weak centrosymmetric
(e.g. the Earth's) gravitational field, with position of center of
mass of the atom (i.e., proton) being fixed in the field by some
non-gravitational forces. It is known that, in a weak field
approximation, curved spacetime is characterized by the following
interval \cite{Misner-1,Landau-1}:
\begin{equation}
d s^2 = -\biggl(1 + 2 \frac{\phi}{c^2} \biggl)(cdt)^2 + \biggl(1 -
2 \frac{\phi}{c^2} \biggl) (dx^2 +dy^2+dz^2 ), \ \ \phi = -
\frac{GM}{R} .
\end{equation}
Below we introduce the proper local coordinates,
\begin{equation}
\tilde t = \biggl( 1 + \frac{\phi}{c^2} \biggl)t, \ \ \tilde x =
\biggl( 1 - \frac{\phi}{c^2} \biggl)x, \ \ \tilde y = \biggl( 1 -
\frac{\phi}{c^2} \biggl) y, \ \ \tilde z = \biggl( 1
-\frac{\phi}{c^2} \biggl) z ,
\end{equation}
where interval has the Minkowski's form  \cite{Misner-1,Landau-1},
\begin{equation}
d \tilde s^2 = -(c d \tilde t)^2 + (d \tilde x^2 +d \tilde y^2 + d
\tilde z^2).
\end{equation}
[Here, we stress that, since we are interested in calculations of
some quantum transition amplitudes with the first order accuracy
with respect to the small parameter, $|\frac{\phi}{c^2}| \ll 1$,
we disregard in Eqs.(3)-(5) and therein below all terms of the
order of $\frac{\phi^2}{c^4}$. We pay attention that near the
Earth's surface the above discussed parameter is small and is
equal to $|\frac{\phi}{c^2}| \sim 10^{-9}$.]

Due to the local Lorentz invariance of a spacetime in general
relativity, if we disregard the so-called tidal terms in the
Hamiltonian [i.e., if we don't differentiate the potential
$\phi(R)$], then new wave functions, written in the local proper
coordinates (4) (with fixed proton's position), satisfy at,
$t,\tilde t >0$, the similar Schr\"{o}dinger equation:
\begin{equation}
i \hbar \frac{\partial \tilde \Psi(\tilde r,\tilde t)}{\partial
\tilde t} = \biggl[ m_e c^2 - \frac{\hbar^2}{2m_e} \biggl(
\frac{\partial^2}{
\partial \tilde x^2} + \frac{\partial^2}{ \partial \tilde y^2} +
\frac{\partial^2}{\partial \tilde z^2} \biggl) - \frac{e^2}{\tilde
r} \biggl] \tilde \Psi(\tilde r,\tilde t).
\end{equation}
[Note that it is easy to show that the above disregarded tidal
terms have relative order of $\frac{r_0}{R_0}$, where $r_0$ is the
Bohr radius and $R_0$ is distance between a hydrogen atom and
center of source of gravitational field. Near the Earth's surface
they are very small and are of the relative order of
$\frac{r_0}{R_0} \sim 10^{-17}$.] $\\$ We stress that it is very
important that the wave function (1) is not a solution of the
Schr\"{o}dinger equation (6) anymore and, thus, is not
characterized by definite energy and weight in the gravitational
field (3). Moreover, a general solution of Eq.(6) can be written
in the proper local coordinates in the following way:
\begin{equation}
\tilde \Psi(\tilde r, \tilde t) = \exp \biggl(\frac{-im_e c^2
\tilde t}{\hbar}\biggl) \sum^{\infty}_{n = 1} \tilde a_n
\Psi_n(\tilde r) \exp\biggl(\frac{-i E_n \tilde t}{\hbar}\biggl),
\end{equation}
where the wave functions $\Psi_n(\tilde r)$ are solutions \cite{Park} for the
so-called $nS$ atomic orbitals of a hydrogen atom with energies
$E_n$ and are normalized in the proper local space,
\begin{equation}
\int  \Psi^2_n(\tilde r) \ d^3 \tilde r = 1.
\end{equation}
[Here we stress that, as possible to show, only $1S \rightarrow
nS$ quantum transitions amplitudes are non-zero in a hydrogen atom
in the gravitational field (3), which corresponds only to real
wave functions. Therefore, we keep in Eq.(7) only $nS$ atomic
orbitals and everywhere below disregard difference between
$\Psi_n(r)$ and $\Psi^*_n(r) = \Psi_n(r)$.] $\\$ Note that the
normalized wave function (1) can be rewritten in the proper local
spacetime coordinates (4) in the following way:
\begin{eqnarray}
\Psi_1(\tilde r,\tilde t) = &&\exp \biggl[\frac{-im_e c^2
(1-\frac{\phi}{c^2}) \tilde t}{\hbar}\biggl] \exp \biggl[\frac{-i
E_1 (1-\frac{\phi}{c^2}) \tilde t}{\hbar} \biggl]
\nonumber\\
&&\times \biggl(1+\frac{\phi}{c^2}\biggl)^{3/2} \Psi_1
\biggl[\biggl(1+\frac{\phi}{c^2} \biggl) \tilde r \biggl]
 \ ,
\end{eqnarray}

It is important that the gravitational field (3) can be considered
as a sudden perturbation to the Hamiltonian (2), therefore, at
$t=\tilde t = 0$, the wave functions (7) and (9) have to be equal
to each other:
\begin{equation}
\biggl( 1+\frac{\phi}{c^2} \biggl)^{3/2} \Psi_1\biggl[
\biggl( 1+\frac{\phi}{c^2} \biggl) \tilde r \biggl] =
\sum^{\infty}_{n=1} \tilde a_n \Psi_n (\tilde r).
\end{equation}
From Eq.(10), it directly follows that
\begin{equation}
\tilde a_1 = \biggl(1+\frac{\phi}{c^2}\biggl)^{3/2} \int^{\infty}_0
\Psi_1 \biggl[\biggl(1+\frac{\phi}{c^2}\biggl) \tilde r\biggl] \Psi_1(\tilde r)
\ d^3 \tilde r
\end{equation}
and
\begin{equation}
\tilde a_n = \biggl(1+\frac{\phi}{c^2}\biggl)^{3/2}
\int^{\infty}_0 \Psi_1 \biggl[\biggl(1+\frac{\phi}{c^2}\biggl)
\tilde r\biggl] \Psi_n(\tilde r) \ d^3 \tilde r , \ \ n>1.
\end{equation}

\subsection{3.3. Probabilities and amplitudes}
Below, we calculate quantum mechanical amplitudes (11) and (12) in
a linear approximation with respect to the gravitational
potential,
\begin{equation}
\tilde a_1 = 1 + O \biggl( \frac{\phi^2}{c^4} \biggl),
\end{equation}
and
\begin{equation}
\tilde a_n = \biggl(\frac{\phi}{c^2} \biggl) \int^{\infty}_0
\biggl[\frac{d \Psi_1(\tilde r)}{d \tilde r} \biggl] \tilde r \Psi_n(\tilde r)
d^3 \tilde r, \ \ n>1.
\end{equation}
We stress that the wave function (7) is a series of wave
functions, which have definite weights in the gravitational field
(3). This means that they are characterized by the following
definite electron passive gravitational masses,
\begin{equation}
m_n = m_e + \frac{E_n}{c^2}.
\end{equation}
In accordance with the most general properties of quantum
mechanics, this means that, if we do a measurement of
gravitational mass for wave function (1) and (9), we obtain
quantum values (15) with the probabilities: $\tilde P_n = |\tilde
a_n|^2$, where $\tilde a_n$ are given by Eqs.(13) and (14).

Let us show that
\begin{equation}
\int^{\infty}_0 \biggl[\frac{d \Psi_1(\tilde r)}{d \tilde r} \biggl]
\tilde r \Psi_n(\tilde r)
d^3 \tilde r = \frac{V_{1n}}{E_n-E_1}, \ \ n>1,
\end{equation}
where $\hat V(\tilde r)$ is the so-called quantum virial operator \cite{Park}:
\begin{equation}
\hat V(r) = - 2 \frac{\hbar^2}{2m_e} \biggl( \frac{\partial^2}{
\partial \tilde x^2} + \frac{\partial^2}{ \partial \tilde y^2} +
\frac{\partial^2}{ \partial \tilde z^2} \biggl) - \frac{e^2}{\tilde r},
\end{equation}
and
\begin{equation}
V_{1,n}=\int^{\infty}_0 \Psi_1(\tilde r) \hat V(\tilde r) \Psi_n(\tilde r)
d^3 \tilde r.
\end{equation}
To this end, we rewrite the Schr\"{o}dinger equation in
gravitational field (6) in terms of the initial coordinates
$(x,y,z)$:
\begin{eqnarray}
(m_e c^2 + E_1)
\Psi_1\biggl[\biggl(1-\frac{\phi}{c^2}\biggl)r\biggl] = \biggl[
m_e c^2&&- \frac{1}{(1-\phi/c^2)^2} \frac{\hbar^2}{2m} \biggl(
\frac{\partial^2}{
\partial x^2} + \frac{\partial^2}{ \partial y^2} +
\frac{\partial^2}{\partial z^2} \biggl)
\nonumber\\
&&- \frac{1}{(1-\phi/c^2)}
\frac{e^2}{r} \biggl]
\Psi_1\biggl[\biggl(1-\frac{\phi}{c^2}\biggl)r\biggl] .
\end{eqnarray}
Then, keeping as usual only terms of the first order with respect
to the small parameter $|\frac{\phi}{c^2}| \ll 1$, we obtain:
\begin{eqnarray}
E_1 \Psi_1(r) - \frac{\phi}{c^2} E_1 r \biggl[\frac{d \Psi_1(r)}{dr} \biggl]=
&&\biggl[ - \frac{\hbar^2}{2m_e} \biggl( \frac{\partial^2}{
\partial x^2} + \frac{\partial^2}{ \partial y^2} +
\frac{\partial^2}{\partial z^2} \biggl) - \frac{e^2}{r} + \frac{\phi}{c^2} \hat
V(r) \biggl]
\nonumber\\
&&\times\biggl[ \Psi_1(r) - \frac{\phi}{c^2} r \biggl[\frac{d
\Psi_1(r)}{dr} \biggl]\biggl],
\end{eqnarray}
and as a result
\begin{eqnarray}
- E_1 r \biggl[\frac{d \Psi_1(r)}{dr} \biggl] = &&\biggl[ -
\frac{\hbar^2}{2m_e} \biggl( \frac{\partial^2}{
\partial x^2} + \frac{\partial^2}{ \partial y^2} +
\frac{\partial^2}{\partial z^2} \biggl) - \frac{e^2}{r} \biggl]
\nonumber\\
&&\times\biggl[- r \frac{d \Psi_1(r)}{dr} \biggl] +\hat V(r) \Psi_1 (r).
\end{eqnarray}
Let us multiply Eq.(21) on $\Psi_1(r)$ and integrate over space,
\begin{eqnarray}
- E_1 \int^{\infty}_{0} \Psi_n(r) r \biggl[\frac{d \Psi_1(r)}{dr}
\biggl] d^3r&& = \int^{\infty}_{0} \Psi_n(r) \biggl[ -
\frac{\hbar^2}{2m_e} \biggl( \frac{\partial^2}{
\partial x^2} + \frac{\partial^2}{ \partial y^2} +
\frac{\partial^2}{\partial z^2} \biggl) - \frac{e^2}{r} \biggl]
\nonumber\\
&&\times \biggl[- r \frac{d \Psi_1(r)}{dr} \biggl]d^3r +\int^{\infty}_{0}
\Psi_n(r)\hat V(r) \Psi_1 (r)d^3r.
\end{eqnarray}
Taking into account that the Hamiltonian operator is the Hermitian
one, we rewrite Eq.(22) as
\begin{eqnarray}
E_1 \int^{\infty}_{0} \Psi_n(r) r \biggl[\frac{d \Psi_1(r)}{dr}
\biggl] d^3r &&= E_n \int^{\infty}_{0} \Psi_n(r) r \biggl[\frac{d
\Psi_1(r)}{dr} \biggl] d^3r
\nonumber\\
&&- \int^{\infty}_{0} \Psi_1(r)\hat V(r)
\Psi_n(r)d^3r.
\end{eqnarray}
Then, Eqs.(16)-(18) directly follow from Eq.(23).

As a result, the calculated amplitudes (14) and the corresponding
probabilities for $n \neq 1$ can be rewritten as functions of
matrix elements (18) of the virial operator (17),
\begin{equation}
\tilde a_n = \biggl( \frac{\phi}{c^2} \biggl) \frac{V_{1,n}}{E_n-E_1}
\end{equation}
and
\begin{equation}
\tilde P_n = |\tilde a_n|^2 = \biggl( \frac{\phi}{c^2} \biggl)^2 \biggl(
\frac{V_{1,n}}{E_n-E_1} \biggl)^2.
\end{equation}
Note that near the Earth's surface, where $\frac{\phi^2}{c^4}
\approx 0.49 \times 10^{-18}$, the probability for $n=2$ in a
hydrogen atom can be calculated as
\begin{equation}
\tilde P_2 = |\tilde a_2|^2 = 1.5 \times 10^{-19},
\end{equation}
where
\begin{equation}
 \frac{V_{1,2}}{E_2-E_1} = 0.56.
\end{equation}
It is important that non-zero matrix elements (18) of the virial
operator (17) for $n \neq 1$ are also responsible for breakdown
of the equivalence between active gravitational mass and energy for
a quantum body with internal degrees of freedom \cite{Lebed-1}.

\section{4. Inequivalence of passive gravitational mass and energy
at microscopic level (method-2)}

\subsection{4.1. Schr\"{o}dinger equation with a definite energy in
the absence of gravitational field}

As in the previous Section, at $t<0$, gravitational field is zero
and electron occupies ground state in a hydrogen atom,
characterizing by the wave function (1). As we have already
discussed, the wave function (1) corresponds to the $1S$ electron
orbital and is known to be a ground state solution of Eq.(2).

\subsection{4.2. Schr\"{o}dinger equation in the presence of gravitational
field}

Let us consider the same Gedanken experiment as in Sec. 3. We
switch on the weak gravitational field (3) and obtain Eq.(6) for
the wave functions in the proper local spacetime coordinates (4).
But, in this Section, we rewrite Eq.(6) in the initial spacetime
coordinates, $(t,x,y,z)$,
\begin{eqnarray}
&&i \hbar \frac{\partial \Psi({\bf r}, t)}{\partial t} = \biggl\{
 \biggl[ m_e c^2 -
\frac{\hbar^2}{2m_e} \biggl( \frac{\partial^2}{
\partial x^2} + \frac{\partial^2}{ \partial y^2} +
\frac{\partial^2}{\partial z^2} \biggl) - \frac{e^2}{r}\biggl]
\nonumber\\
&&+ \biggl(\frac{\phi}{c^2}\biggl) \biggl[ m_e c^2 -
\frac{\hbar^2}{2m_e} \biggl( \frac{\partial^2}{
\partial x^2} + \frac{\partial^2}{ \partial y^2} +
\frac{\partial^2}{\partial z^2} \biggl) - \frac{e^2}{r} + \hat
V({\bf r}) \biggl] \biggl\} \Psi({\bf r},t),
\end{eqnarray}
where the virial operator \cite{Park}, $\hat V(r)$, is equal to
(17). From Eq.(28), it directly follows that the external
gravitational field (3) is coupled not only to Hamiltonian (2) but
also to the virial operator (17). It is important that the virial
term (17) does not commute with the Hamiltonian (2), therefore, it
breaks the equivalence of the passive gravitational mass and
energy for electron in a hydrogen atom.

\subsubsection{4.2.1. More general Lagrangian}

Here, we derive Hamiltonian (28) from more general Lagrangian. Let
us consider the Lagrangian of a three body system: a hydrogen atom
and the Earth in inertial coordinate system, treating gravitation
(3) as a small perturbation in the Minkowski's spacetime. In this
case, we can make use of the results of Ref.[7], where the
corresponding n-body Lagrangian is calculated as a sum of the
following four terms:
\begin{equation}
L = L_{kin} + L_{em} + L_{G} + L_{e,G},
\end{equation}
where $L_{kin}$, $L_{em}$, $L_G$, and $L_{e,G}$ are kinetic,
electromagnetic, gravitational and electric-gravitational parts of
the Lagrangian, respectively. We recall that, in our
approximation, we keep in the Lagrangian and Hamiltonian only
terms of the order of $(v/c)^2$ and $|\phi|/c^2$ as well as keep
only classical kinetic and the Coulomb electrostatic potential
energies couplings with external gravitational field. It is
possible to show that, in our case, different contributions to the
Lagrangian (29) can be simplified:

\begin{equation}
L_{kin} + L_{em} = - Mc^2 - m_pc^2 - m_ec^2 + m_e\frac{{\bf
v}^2}{2} + \frac{e^2}{r} \ ,
\end{equation}

\begin{equation}
L_{G} = G \frac{m_p M}{R} + G \frac{m_e M}{R} + \frac{3}{2} G
\frac{m_e M}{R} \frac{{\bf v^2}}{c^2} \ ,
\end{equation}

\begin{equation}
L_{e,G} = - 2 G \frac{M}{Rc^2} \frac{e^2}{r} \ ,
\end{equation}
where, as usual, we use the inequality $m_p \gg m_e$, with $m_p$
being the bare proton mass.

If we keep only those terms in the Lagrangian, which are related
to electron motion (as usual, proton is supposed to be supported
by some non-gravitational forces in the gravitational field), then
we can write the Lagrangian (30)-(32) in the following form:
\begin{equation}
L = -m_e c^2 + m_e \frac{{\bf v}^2}{2} + \frac{e^2}{r} - \frac{\phi(R)}{c^2}
\biggl[ m_e + 3 m_e \frac{{\bf v}^2}{2} - 2 \frac{e^2}{r} \biggl]
\ , \ \ \ \phi(R) = - G \frac{M}{R}.
\end{equation}
 It is easy to show that the
corresponding electron Hamiltonian is

\begin{eqnarray}
&&\hat H = \biggl\{
 \biggl[ m_e c^2 -
\frac{\hbar^2}{2m_e} \biggl( \frac{\partial^2}{
\partial x^2} + \frac{\partial^2}{ \partial y^2} +
\frac{\partial^2}{\partial z^2} \biggl) - \frac{e^2}{r}\biggl]
\nonumber\\
&&+ \biggl(\frac{\phi}{c^2}\biggl) \biggl[ m_e c^2 -
\frac{\hbar^2}{2m_e} \biggl( \frac{\partial^2}{
\partial x^2} + \frac{\partial^2}{ \partial y^2} +
\frac{\partial^2}{\partial z^2} \biggl) - \frac{e^2}{r} + \hat
V({\bf r}) \biggl] \biggl\}.
\end{eqnarray}

Note that Eq.(34) exactly coincides with electron Hamiltonian
(28), obtained by us in the previous Subsection.

\subsubsection{4.2.2. More general Hamiltonian}

Let us derive the Hamiltonian (28),(34) from more general
arguments. The so-called gravitational Stark effect (i.e., the
mixing effect between even and odd wave functions in a hydrogen
atom in gravitational field) was studied in Ref.[13] in the weak
external gravitational field (3). Note that the corresponding
Hamiltonian was derived in $1/c^2$ approximation and a possibility
of center of mass of the atom motion was taken into account. The
main peculiarity of the calculations in the above-mention paper
was the fact that not only terms of the order of $\phi/c^2$ were
calculated, as in our case, but also terms of the order of
$\phi'/c^2$. Here, we use a symbolic notation $\phi'$ for the
first derivatives of gravitational potential. In accordance with
the existing tradition, we refer to the latter terms as to the
tidal ones. Note that the Hamiltonian (3.24) was obtained in
Ref.[13] directly from the Dirac equation in a curved spacetime of
general relativity. As shown in
 Ref.[13], it can be rewritten for the corresponding
Schr\"{o}dinger equation as a sum of the four terms:
\begin{equation}
\hat H (\hat {\bf P}, \hat {\bf p}, \tilde {\bf R},r)= \hat H_0
(\hat {\bf P}, \hat {\bf p}, r) + \hat H_1 (\hat {\bf P}, \hat
{\bf p}, \tilde {\bf R},r) + \hat H_2 (\hat {\bf p}, {\bf r}) +
\hat H_3 (\hat {\bf P}, \hat {\bf p},\tilde {\bf R},r) ,
\end{equation}

\begin{equation}
\hat H_0 (\hat {\bf P}, \hat {\bf p}, r) = m_e c^2 + m_p c^2 +
\biggl[\frac{\hat {\bf P}^2}{2(m_e + m_p)} + \frac{\hat {\bf
p}^2}{2 \mu} \biggl] - \frac{e^2}{r} ,
\end{equation}

\begin{equation}
\hat H_1 (\hat {\bf P}, \hat {\bf p}, \tilde {\bf R}, r) =
\biggl\{ m_e c^2 + m_p c^2 +  \biggl[3 \frac{\hat {\bf P}^2}{2(m_e
+ m_p)} + 3 \frac{\hat {\bf p}^2}{2 \mu} - 2 \frac{e^2}{r}
\biggl]\biggl\} \biggl( \frac{\phi  - {\bf g}\tilde {\bf R}}{c^2}
\biggl),
\end{equation}

\begin{eqnarray}
&&\hat H_2 (\hat {\bf p}, {\bf r}) = \frac{1}{c^2}
\biggl(\frac{1}{m_e}-\frac{1}{m_p} \biggl)[-({\bf g}{\bf r}) \hat
{\bf p}^2 + i \hbar {\bf g} \hat {\bf p}]
\nonumber\\
&&+\frac{1}{c^2} {\bf g} \biggl(\frac{\hat {\bf s_e}}{m_e} -
\frac{\hat {\bf s_p}}{m_p} \biggl) \times \hat {\bf p} + \frac{e^2
(m_p-m_e)}{2(m_e+m_p)c^2} \frac{{\bf g}{\bf r}}{r},
\end{eqnarray}

\begin{eqnarray}
\hat H_3 (\hat {\bf P}, \hat {\bf p}, \tilde {\bf R}, r) &&=
\frac{3}{2}\frac{i \hbar {\bf g}{\bf P}}{(m_e+m_p)c^2}
+\frac{3}{2} \frac{{\bf g}{\bf(s_e+s_p)}\times {\bf
P}}{(m_e+m_p)c^2}
\nonumber\\
&&- \frac{({\bf g}{\bf r})({\bf
P}{\bf p})+({\bf P}{\bf r})({\bf g}{\bf p})-i\hbar {\bf g}{\bf
P}}{(m_e+m_p)c^2},
\end{eqnarray}
where ${\bf g}=-G \frac{M}{R^3} {\bf R}$. Note that we use the
following notations in Eqs.(35)-(39): $\tilde {\bf R}$ and ${\bf
P}$ stand for coordinate and momentum of a hydrogen atom center of
mass, respectively; whereas, ${\bf r}$ and ${\bf p}$ stand for
relative electron coordinate and momentum in center of mass
coordinate system; $\mu = m_e m_p /(m_e + m_p)$ is the so-called
reduced electron mass. We point out that $\hat H_0 (\hat {\bf P},
\hat {\bf p}, r)$ is the Hamiltonian of a hydrogen atom in the
absence of the field. It is important that the Hamiltonian $\hat
H_1 (\hat {\bf P}, \hat {\bf p}, \tilde {\bf R}, r)$ describes
couplings not only of the bare electron and proton masses with the
gravitational field (3) but also couplings of electron kinetic and
potential energies with the field. And finally, the Hamiltonians
$\hat H_2 (\hat {\bf p}, {\bf r})$ and $\hat H_3 (\hat {\bf P},
\hat {\bf p}, \tilde {\bf R}, r)$ describe only the tidal effects.

Let us strictly derive the Hamiltonian (28),(34), which has
already been semi-quantitatively derived, from the more general
Hamiltonian (35)-(39). As was already mentioned, we use the
approximation, where $m_p \gg m_e$, and, therefore, $\mu = m_e$.
In particular, this allows us to consider proton as a heavy
classical particle. We recall that we need to derive the
Hamiltonian of the atom, whose center of mass is at rest with
respect to the Earth. Thus, we can omit center of mass kinetic
energy and center of mass momentum. As a result, the first two
contributions to electron part of the total Hamiltonian (35)-(39)
can be written in the following way:
\begin{equation}
\hat H_0 (\hat {\bf p}, r) = m_e c^2 + \frac{\hat {\bf p}^2}{2m_e}
- \frac{e^2}{r}
\end{equation}
and
\begin{equation}
\hat H_1 (\hat {\bf p}, r) =  \biggl\{ m_e c^2 + \biggl[3
\frac{\hat {\bf p}^2}{2 m_e} - 2 \frac{e^2}{r} \biggl]\biggl\}
\biggl( \frac{\phi}{c^2} \biggl),
\end{equation}
where we place center of mass of the atom at point $\tilde {\bf R}
= 0$. Now, let us study the first tidal term (38) in the total
Hamiltonian (35). At first, we pay attention that $|{\bf g}|
\simeq |\phi|/R_0$. Then, as well known, in a hydrogen atom $|{\bf
r}| \sim \hbar / |{\bf p}| \sim r_B$ and ${\bf p}^2/(2m_e) \sim
e^2/r_B$. These values allow us to evaluate the first tidal term
(38) in the Hamiltonian (35) as $H_2 \sim (r_B/R_0) (|\phi|/c^2)
(e^2/r_B) \sim 10^{-17} (|\phi|/c^2) (e^2/r_B)$. Note that this
value is $10^{-17}$ times smaller than $H_1 \sim (|\phi|/c^2)
(e^2/r_B)$ and $10^{-8}$ times smaller than the second correction
with respect to the small parameter $|\phi|/c^2$. Therefore, we
can disregard the contribution (38) to the total Hamiltonian (35).
As to the second tidal term (39) in the total Hamiltonian, we pay
attention that it is exactly zero in the case, where ${\bf P}=0$,
considered in this review. Therefore, we can conclude that the
Hamiltonian (40),(41), derived in this Subsubsection, exactly
coincides with that, semi-quantitatively derived by us earlier
[see Eqs.(28),(34)].

\subsection{4.3. Gravitational field as a perturbation to the
Hamiltonian}

It is important that the gravitational field (3), under the
condition of our Gedanken experiment, can be considered as the
following sudden perturbation, $\hat U_1({\bf r},t)$, to the
Hamiltonian (2) in the absence of gravitational field:
\begin{equation}
\hat U_1({\bf r},t) = \biggl(\frac{\phi}{c^2}\biggl) \biggl[ m_e
c^2 - \frac{\hbar^2}{2m_e} \biggl( \frac{\partial^2}{
\partial x^2} + \frac{\partial^2}{ \partial y^2} +
\frac{\partial^2}{\partial z^2} \biggl) - \frac{e^2}{r} + \hat
V({\bf r}) \biggl] \Theta(t) ,
\end{equation}
where $\Theta(t)$ is the so called step function. Then, a general
solution of Eq.(28) can be written in the following way:
\begin{eqnarray}
&&\Psi(r, t) = \exp \biggl(\frac{-i \tilde m_e c^2
t}{\hbar}\biggl) \Psi^1_1(r) \exp\biggl(\frac{-i \tilde E_1
t}{\hbar}\biggl)
\nonumber\\
&&+\exp \biggl(\frac{-im_e c^2 t}{\hbar}\biggl) \sum^{\infty}_{n >
1} a_n \Psi_n(r) \exp\biggl(\frac{-i E_n t}{\hbar}\biggl) \ ,
\end{eqnarray}
where the wave functions $\Psi_n(r)$ are solutions for the $nS$
orbitals in a hydrogen atom and are normalized,
\begin{equation}
\int  [\Psi^1_1(r)]^2 \ d^3 r = 1; \ \ \int  [\tilde \Psi_n(r)]^2
\ d^3 r = 1, \ \ n>1.
\end{equation}
[It is easy to show that perturbation (42) can results only in
non-zero quantum transitions between $1S$ and $nS$ electron
orbitals, therefore, we keep in Eq.(43) only $\Psi_n(r)$ wave
functions, which are real.]

According with the standard time-dependent perturbation theory
\cite{Park}, the corrected wave-function of ground state,
$\Psi^1_1(r)$, as well as the corrections to mass and energy of
ground state in Eq.(43) can be written as:
\begin{eqnarray}
&&\Psi^1_1(r) = \Psi_1(r) + \biggl(\frac{\phi}{c^2} \biggl)
\sum^{\infty}_{n> 1} \frac{V_{n,1}}{E_1-E_n} \Psi_n(r),
\nonumber\\
&&\tilde m_e = \biggl(1 + \frac{\phi}{c^2} \biggl) m_e, \ \ \tilde
E_1 = \biggl(1 + \frac{\phi}{c^2} \biggl) E_1,
\end{eqnarray}
where $V_{n,1}$ is matrix element of the virial operator (17):
\begin{equation}
 V_{n,1} = \int \Psi_n(r)
\biggl[ - 2 \frac{\hbar^2}{2m} \biggl( \frac{\partial^2}{
\partial x^2} + \frac{\partial^2}{ \partial y^2} +
\frac{\partial^2}{\partial z^2} \biggl)- \frac{e^2}{r} \biggl]
\Psi_1(r) d^3 {\bf r} .
\end{equation}
Note that the very last term in Eq.(45) corresponds to the
so-called red shift in gravitational field. It is due to the
expected contribution to passive gravitational mass from electron
binding energy in the atom. As to the coefficients $a_n$ with $n
\neq 1$ in Eq.(43), they can be also written in terms of the
virial operator matrix elements,
\begin{equation}
a_n = - \biggl( \frac{\phi}{c^2} \biggl) \biggl(
\frac{V_{n,1}}{E_1-E_n} \biggl),
\end{equation}
and coincides with Eq.(24). Note that the wave function (43)-(47),
which corresponds to electron ground energy level in the presence
of the gravitational field (3) (i.e., at $t > 0$), is a series of
eigenfunctions of electron energy operator, taken in the absence
of the field. Therefore, if we measure energy, in electron quantum
state (43)-(47), we obtain the following quantized values for
electron gravitational mass:
\begin{equation}
m_n = m_e + \frac{E_n}{c^2},
\end{equation}
where we omit the red shift effect. From Eqs.(43)-(48), we can
state that the expected Einstein's equation, $m = m_e +
\frac{E_1}{c^2}$, survives in our case with probability close to
1, whereas with the following small probabilities,
\begin{equation}
P_n = |a_n|^2 = \biggl(\frac{\phi}{c^2}\biggl)^2
\frac{V^2_{n,1}}{(E_n -E_1)^2} , \ \ \ n \neq 1 ,
\end{equation}
it is broken. The reason for this breakdown is that, the virial
term (17) does not commute with the Hamiltonian (2) in the absence
of gravitational field. As a result, electron wave functions with
definite passive gravitational masses are not characterized by
definite energies in the absence of gravitational field. It is
important that our current results coincide with that obtained in
Sec.3 by different method.

\subsection{4.4. Experimental aspects}

Here, let us describe another Gedanken experiment, where
gravitational field is adiabatically switched on. To this end, we
consider wave function (1) to be valid at $t \rightarrow -\infty$
and apply the following perturbation, due to the gravitational
field (3), for the Hamiltonian (2):
\begin{equation}
\hat U_2({\bf r},t) = \biggl(\frac{\phi}{c^2}\biggl) \biggl[ m_e
c^2 - \frac{\hbar^2}{2m_e} \biggl( \frac{\partial^2}{
\partial x^2} + \frac{\partial^2}{ \partial y^2} +
\frac{\partial^2}{\partial z^2} \biggl) - \frac{e^2}{r} + \hat
V({\bf r}) \biggl] \ \exp(\lambda t), \ \lambda \rightarrow 0.
\end{equation}
Then, at $t \simeq 0$ (i.e., in the presence of the field), the
electron wave function can be written as
\begin{eqnarray}
&&\Psi(r, t) = \exp \biggl(\frac{-i \tilde m_e c^2
t}{\hbar}\biggl) \Psi^1_1(r) \exp\biggl(\frac{-i \tilde E_1
t}{\hbar}\biggl)
\nonumber\\
&&+\exp \biggl(\frac{-im_e c^2 t}{\hbar}\biggl) \sum^{\infty}_{n >
1} a_n \Psi_n(r) \exp\biggl(\frac{-i E_n t}{\hbar}\biggl) \ .
\end{eqnarray}
Application of the standard time-dependent perturbation theory
\cite{Park} in the case of adiabatic switching on gravitational
field results in:
\begin{eqnarray}
&&\Psi^1_1(r) = \Psi_1(r) + \biggl(\frac{\phi}{c^2} \biggl)
\sum^{\infty}_{n> 1} \frac{V_{n,1}}{E_1-E_n} \Psi_n(r),
\nonumber\\
&&\tilde m_e = \biggl(1 + \frac{\phi}{c^2} \biggl) m_e, \ \ \tilde
E_1 = \biggl(1 + \frac{\phi}{c^2} \biggl) E_1,
\end{eqnarray}
and
\begin{equation}
a_n=0, \ \ \ P_n =0.
\end{equation}

Thus, in adiabatic limit, the phenomenon of quantization of
passive gravitational mass (15),(48) disappears. This means that
the possible experimental observation of the above mentioned
phenomenon has to be done in quickly changing gravitational field.
It is important that step-like function, $\Theta(t)$, which was
used to derive Eq.(48), does not mean motion of a source of
gravity with velocity higher than the speed of light. We can use
step-like function if significant change of gravitational field
happens quicker than the characteristic period of quasiclassical
rotation of electron in a hydrogen atom. In the case under
consideration, we need the time about $\delta t \leq t_0 =\frac{2
\pi \hbar}{E_2 -E_1} \sim 10^{-15}s$. Of course, there exist much
more convenient quantum systems with higher values of the
parameter $t_0$, where the above discussed phenomenon could be
observed. We recall that all excited energy levels are
quasistationary and, thus, decay with time by emitting photons.
Therefore, it is much more efficient to detect emitted photons
than to directly measure a weight. As to the relatively small
probabilities (24) of the mass quantization, it is not too small
and can be compensated by large value of the Avogadro number, $N_A
\approx 6 \times 10^{23}$. In other words, for macroscopic number
of the atoms, we may have large number of emitted photons. For
instance, the number of excited electrons (i.e., emitted photons)
for 1000 moles of the atoms is estimated as
\begin{equation}
N_n = 2.95 \times 10^{8} \times \biggl( \frac{V_{n,1}}{E_n-E_1}
\biggl)^2 , \ \ \ N_2 = 0.9 \times 10^8.
\end{equation}

\section{5. Inequivalence between passive gravitational mass and energy
at macroscopic level} In Sec. 5, we perform our Gedanken
experiment, where we switch on the gravitational field (3) for $t
> 0$, using the gravitational field as a sudden perturbation (42). We
consider two different cases: macroscopic ensemble of stationary
quantum states and macroscopic ensemble of coherent superpositions
of stationary quantum states. In this section we disregard small
probabilities of the order of $\frac{\phi^2}{c^2}$ [see Eqs.(25)
and (49)] and, thus, ignore mass quantization phenomenon.

\subsection{5.1. Equivalence between passive gravitational mass and energy
of stationary quantum states}

Suppose that, at $t < 0$, there is no gravitational field and we
have macroscopic ensemble of hydrogen atoms with electrons being
in their ground states (1). At $t > 0$, we perform our Gedanken
experiment: we switch on gravitational field, which is treated as
the perturbation (42) in inertial system. Let us for the moment
consider one atom. At $t > 0$, general solution for electron wave
function is
\begin{eqnarray}
&&\Psi(r, t) = \exp \biggl(\frac{-i \tilde m_e c^2
t}{\hbar}\biggl) \Psi^1_1(r) \exp\biggl(\frac{-i \tilde E_1
t}{\hbar}\biggl)
\nonumber\\
&&+\exp \biggl(\frac{-im_e c^2 t}{\hbar}\biggl) \sum^{\infty}_{n >
1} a_n \Psi_n(r) \exp\biggl(\frac{-i E_n t}{\hbar}\biggl) \ ,
\end{eqnarray}
If we disregard small probabilities $a_n$ for $n>1$, which were
considered in Secs. 3 and 4, we can rewrite the Eq.(55) as
\begin{equation}
\Psi(r, t) = \exp \biggl(\frac{-i \tilde m_e c^2
t}{\hbar}\biggl) \Psi^1_1(r) \exp\biggl(\frac{-i \tilde E_1
t}{\hbar}\biggl) .
\end{equation}
In accordance with the quantum perturbation theory \cite{Park}, first
order correction to energy of wave function (56) can be written as:
\begin{equation}
\tilde m_e = \biggl(1+\frac{\phi}{c^2}\biggl) m_e, \ \ \ \tilde E_1
= \biggl(1+\frac{\phi}{c^2}\biggl) E_1 ,
\end{equation}
which is well known red shift \cite{Misner-1}. It is important
that, in Eq.(57), there is no correction due to the quantum virial
term (42),(46). The virial term correction is zero due the
so-called quantum virial theorem \cite{Park}, which claims that
for any value of $n$, including $n=1$:
\begin{equation}
V_{n,n} = \int \Psi_n(r) \hat V({\bf r}) \Psi_n(r)d^3{\bf r} =0.
\end{equation}
Eq.(57) directly demonstrates the equivalence of gravitational mass and
energy at macroscopic level.

\subsection{5.2. Inequivalence between passive gravitational mass and
energy for macroscopic coherent ensemble of superpositions of
stationary quantum states}
Suppose that, in the absence of
gravitational field (i.e., at $t<0$), we have macroscopic ensemble
of coherent superpositions of two wave functions, corresponding to
ground state wave function, $\Psi_1(r)$, and first excited energy
level wave function, $\Psi_2(r)$, in a hydrogen atom:
\begin{equation}
\Psi(r,t) = \frac{1}{\sqrt{2}}\exp \biggl( \frac{-i m_e c^2 t}{
\hbar} \biggl)\biggl[ \exp \biggl(\frac{-i E_1 t}{ \hbar} \biggl)
\Psi_1(r) + \exp \biggl(\frac{-i E_2 t}{ \hbar} \biggl)
\Psi_2(r)\biggl] .
\end{equation}
Coherent ensemble of such wave functions, where the difference
between phases of functions $\Psi_1(r)$ and $\Psi_2(r)$ is fixed,
is possible to create by using lasers. We perform the same
Gedanken experiment and, therefore, saddenly switch on the
gravitational field (3) at $t>0$ [see the corresponding
perturbation (42) to the Hamiltonian (2)]:
\begin{equation}
U_1({\bf r},t) = \frac{\phi}{c^2}[m_e c^2 + \hat H_0({\bf r})
+\hat V({\bf r})] \Theta(t).
\end{equation}
Then, if we disregard small probabilities of the order of
$\frac{\phi^2}{c^4}$ [see Eq.(25)] and, thus, don't take into
account the mass quantization phenomenon (15), we can consider
wave function (59) as wave function of two level system and can
use the corresponding variant of the time-dependent perturbation
theory. In accordance with this theory \cite{Park}, the wave
functions in the gravitational field (3) can be written as:
\begin{eqnarray}
\Psi^1(r,t) = \exp \biggl( \frac{-i m_e c^2 t}{ \hbar}
\biggl)\biggl[ &&\exp \biggl(\frac{-i E_1 t}{ \hbar} \biggl)
a_1(t) \Psi_1(r)
\nonumber\\
&&+ \exp \biggl(\frac{-i E_2 t}{
\hbar} \biggl) a_2(t) \Psi_2(r)\biggl] .
\end{eqnarray}
Using the results of the time-dependent perturbation theory, it is
possible to find equations to determine the functions $a_1(t)$ and
$a_2(t)$:
\begin{eqnarray}
&&\frac{da_1(t)}{dt} = - i \ U_{11}(t)\ a_1(t) -i \ U_{12}(t) \exp
\biggl[-i\frac{(E_2-E_1) t}{\hbar} \biggl] \ a_2(t),
\nonumber\\
&&\frac{da_2(t)}{dt} = - i \ U_{22}(t)\ a_2(t) -i \ U_{21}(t) \exp
\biggl[-i\frac{(E_1-E_2) t}{\hbar} \biggl] \ a_1(t),
\end{eqnarray}
where
\begin{eqnarray}
&&U_{11}(t)= \Theta(t)\frac{\phi}{c^2} \int \Psi^*_1(r)[m_ec^2 +
\hat H({\bf r}) + \hat V({\bf r})] \Psi_1(r) d^3{\bf
r}=\Theta(t)\frac{\phi}{c^2}(m_ec^2+E_1),
\nonumber\\
&&U_{12}(t)= \Theta(t)\frac{\phi}{c^2} \int \Psi^*_1(r)[m_ec^2 +
\hat H({\bf r}) + \hat V({\bf r})] \Psi_2(r) d^3{\bf
r}=\Theta(t)\frac{\phi}{c^2} V_{12},
\nonumber\\
&&U_{22}(t)= \Theta(t)\frac{\phi}{c^2} \int \Psi^*_2(r)[m_ec^2 +
\hat H({\bf r}) +\hat V({\bf r})] \Psi_2(r) d^3{\bf
r}=\Theta(t)\frac{\phi}{c^2}(m_ec^2+E_2),
\nonumber\\
&&U_{21}(t)= \Theta(t)\frac{\phi}{c^2} \int \Psi^*_2(r)[m_ec^2 +
\hat H({\bf r}) + \hat V({\bf r})] \Psi_1(r) d^3{\bf
r}=\Theta(t)\frac{\phi}{c^2} V_{21},
\end{eqnarray}
where $V_{ij}$ are matrix elements of the virial operator (17).
After substitution of Eqs.(63) in Eqs.(62), it possible to find
that the function (59) is
\begin{equation}
\Psi^1(r,t) = \exp \biggl( \frac{-i m_e c^2 t}{ \hbar}
\biggl)\biggl[ \Psi^1_{1}(r,t) + \Psi^1_{2}(r,t)\biggl],
\end{equation}
where
\begin{eqnarray}
\Psi^1_{1}(r,t) = \frac{1}{\sqrt{2}} \exp \biggl[ -i \frac{(m_e
c^2 + E_1) \phi t}{c^2\hbar}&&\biggl]\exp \biggl(-i \frac{E_1
t}{\hbar} \bigg) \biggl[1 -\frac{\phi V_{12}}{c^2(E_2-E_1)}
\biggl] \Psi_1(r)
\nonumber\\
&&+\frac{1}{\sqrt{2}} \exp \biggl(-i \frac{E_2 t}{\hbar}
\bigg)\frac{\phi V_{12}}{c^2(E_2-E_1)}  \Psi_1(r)
\end{eqnarray}
and
\begin{eqnarray}
\Psi^1_{2}(r,t) = \frac{1}{\sqrt{2}} \exp \biggl[ -i \frac{(m_e
c^2 + E_2) \phi t}{c^2\hbar}&&\biggl] \exp \biggl(-i \frac{E_2
t}{\hbar} \bigg) \biggl[1 -\frac{\phi V_{21}}{c^2(E_1-E_2)}
\biggl] \Psi_2(r)
\nonumber\\
&&+\frac{1}{\sqrt{2}} \exp \biggl(-i \frac{E_1 t}{\hbar}
\bigg)\frac{\phi V_{21}}{c^2(E_1-E_2)}  \Psi_2(r).
\end{eqnarray}

It is possible to demonstrate that with accuracy to the first
order of the small parameter, $\frac{|\phi|}{c^2} \ll 1$, the wave
function (64)-(66) can be written as
\begin{eqnarray}
\Psi^1(r,t) = \frac{1}{\sqrt{2}} \exp \biggl[ -i \frac{(m_e c^2 +
E_1) (1+\phi) t}{c^2\hbar}\biggl] \biggl\{&&\biggl[1 -\frac{\phi
V_{12}}{c^2(E_2-E_1)} \biggl] \Psi_1(r)
\nonumber\\
&&+\frac{\phi V_{21}}{c^2(E_1-E_2)}  \Psi_2(r) \biggl\}
\nonumber\\
+\frac{1}{\sqrt{2}} \exp \biggl[ -i \frac{(m_e c^2 + E_2) (1+\phi)
t}{c^2\hbar}\biggl] \biggl\{&&\biggl[1 -\frac{\phi
V_{21}}{c^2(E_1-E_2)} \biggl] \Psi_2(r)
\nonumber\\
&&+\frac{\phi V_{12}}{c^2(E_2-E_1)}  \Psi_1(r) \biggl\},
\end{eqnarray}
where the wave function is normalized with the same accuracy:
\begin{equation}
\int [\Psi^1(r,t)]^*\Psi^1(r,t) d^3r = 1 + O
\biggl(\frac{\phi^2}{c^4} \biggl).
\end{equation}
Taking into account that we consider macroscopic coherent ensemble
of superposition of quantum states (59),(67), it is easy to
calculate the expectation value of energy per one electron in
gravitational field for wave function (67):
\begin{equation}
<E>= m_ec^2 \biggl(1 +\frac{\phi}{c^2}\biggl) + \frac{(E_1+E_2)}{2
c^2}\biggl(1 +\frac{\phi}{c^2}\biggl)+ V_{12} \frac{\phi}{c^2}.
\end{equation}
Note that the first term and the second one are expected. On the
other hand, the last term contains contribution to the weight of
macroscopic coherent ensemble from the virial term (17) and breaks
the equivalence of passive gravitational mass and energy for
quantum superposition of stationary state.

Note that so far we have considered macroscopic coherent ensemble
of superposition of stationary wave functions, which is
characterized by constant difference of phases, $\alpha =0$,
between the first and the second quantum states. If we introduce
more general macroscopic coherent ensemble,
\begin{eqnarray}
\Psi(r,t) = \frac{1}{\sqrt{2}}\exp \biggl( \frac{-i m_e c^2 t}{
\hbar} \biggl)\biggl[&& \exp \biggl(\frac{-i E_1 t}{ \hbar}
\biggl) \Psi_1(r)
\nonumber\\
&&+ \exp(i\alpha) \exp \biggl(\frac{-i E_2 t}{ \hbar} \biggl)
\Psi_2(r)\biggl] ,
\end{eqnarray}
the expectation value of energy in the gravitational field (3) is
changed:
\begin{equation}
<E>= m_ec^2 \biggl(1 +\frac{\phi}{c^2}\biggl) + \frac{(E_1+E_2)}{2
c^2}\biggl(1 +\frac{\phi}{c^2}\biggl)+ V_{12} \cos \alpha
\frac{\phi}{c^2}.
\end{equation}
It is important that Eqs.(69),(71) directly demonstrate the
breakdown of the equivalence between gravitational mass and energy
at macroscopic level for coherent ensemble of superposition of
stationary quantum state. On the other hand, for the non-coherent
ensembles, phase $\alpha$ is not fixed in Eq.(71) and, thus, the
last virial terms quickly averages to zero.

\section{6. Inequivalence between active gravitational mass and energy
at macroscopic level} In this section, we review our results
\cite{Lebed-1,Lebed-2}, where we showed that active gravitational
mass and energy were inequivalent to each other at macroscopic
level for coherent ensembles of quantum superpositions of
stationary states.

\subsection{6.1. Active gravitational mass in classical physics} Here,
we determine electron active gravitational mass in a classical
model of a hydrogen atom, which takes into account electron
kinetic and potential energies \cite{Nord}. More specifically, we
consider a particle with small bare mass $m_e$, moving in the
Coulomb electrostatic field of a heavy particle with bare mass
$m_p \gg m_e$. Our task is to find gravitational potential at
large distance from the atom, $R \gg r_B$, where $r_B$ is the the
so-called Bohr radius (i.e., effective "size" of a hydrogen atom).
Bellow, we use the so-called weak field gravitational theory
\cite{Misner-1,Nord}, where the post-Newtonian gravitational
potential can be represented as \cite{Lebed-1,Lebed-2}

\begin{equation}
\phi(R,t)=-G \frac{m_p + m_e}{R}- G \int \frac{\Delta
T^{kin}_{\alpha \alpha}(t,{\bf r})+
 \Delta T^{pot}_{\alpha
\alpha}(t,{\bf r})}{c^2R} d^3 {\bf r} ,
\end{equation}
where $\Delta T^{kin}_{\alpha \beta}(t,{\bf r})$ and $\Delta
T^{pot}_{\alpha \beta}(t,{\bf r})$ are contributions to
stress-energy tensor density, $T_{\alpha \beta}(t, {\bf r})$, due
to kinetic and the Coulomb potential energies, respectively. We
point out that, in Eq.(72), we disregard all retardation effects.
Thus, in the above-discussed approximation, electron active
gravitational mass is equal to
\begin{equation}
m^a_e = m_e + \frac{1}{c^2} \int [\Delta T^{kin}_{\alpha
\alpha}(t,{\bf r}) + \Delta T^{pot}_{\alpha \alpha}(t,{\bf r})]
d^3{\bf r}.
\end{equation}

Let us calculate $\Delta T^{kin}_{\alpha \alpha}(t, {\bf r})$,
using the standard expression for stress-energy tensor density of
a moving relativistic point mass \cite{Misner-1,Landau-1}:
\begin{equation}
T^{\alpha \beta}_{kin}({\bf r},t) = \frac{m_e v^{\alpha}(t)
v^{\beta}(t)}{\sqrt{1-v^2(t)/c^2}} \ \delta^3[{\bf r}-{\bf
r}_e(t)],
\end{equation}
where $v^{\alpha}$ is a four-velocity, $\delta^3(...)$ is the
three dimensional Dirac $\delta$-function, and ${\bf r}_e(t)$ is a
three dimensional electron trajectory.

From Eqs.(73),(74), it directly follows that
\begin{equation}
\Delta T^{kin}_{\alpha \alpha}(t) = \int \Delta T^{kin}_{\alpha
\alpha}(t,{\bf r}) d^3{\bf r} = \frac{m_e [c^2
+v^2(t)]}{\sqrt{1-v^2(t)/c^2}} -m_ec^2.
\end{equation}
Note that, although calculations of the contribution from
potential energy to stress energy tensor are more complicated,
they are straightforward and can be done by using the standard
formula for stress energy tensor of electromagnetic field
\cite{Landau-1},
\begin{equation}
T_{em}^{\mu \nu} = \frac{1}{4 \pi} [F^{\mu \alpha} F^{\nu}_{\
\alpha} - \frac{1}{4} \eta^{\mu \nu} F_{\alpha \beta} F^{\alpha
\beta}],
\end{equation}
where $\eta_{\alpha \beta}$ is the Minkowski metric tensor,
$F^{\alpha \beta}$ is the so-called tensor of electromagnetic
field \cite{Landau-1}. In this review, we use approximation, where
we do not take into account magnetic field and keep only the
Coulomb electrostatic field. In this approximation, we can
simplify Eq.(76) and obtain from it the following expression:
\begin{equation}
\Delta T^{pot}_{\alpha \alpha} (t) = \int \Delta T^{pot}_{\alpha
\alpha}(t,{\bf r}) d^3{\bf r} = -2\frac{e^2}{r(t)},
\end{equation}
where $e$ is the electron charge. As directly follows from
Eqs.(75),(77), electron active gravitational mass can be
represented in the following way:
\begin{equation}
m^a_e = \biggl[\frac{m_e c^2}{(1 -v^2/c^2)^{1/2}} - \frac{e^2}{r}
\biggl]/c^2 + \biggl[\frac{m_e v^2}{(1
-v^2/c^2)^{1/2}}-\frac{e^2}{r}\biggl]/c^2.
\end{equation}
We note that the first term in Eq.(78) is the expected one.
Indeed, it is the total energy contribution to the mass, whereas
the second term is the so-called relativistic virial one
\cite{Park}. It is important that it depends on time. Therefore,
in classical physics, active gravitational mass of a composite
body depends on time too. Nevertheless, in this situation, it is
possible to introduce averaged over time electron active
gravitational mass. This procedure results in the expected
equivalence between averaged over time active gravitational mass
and energy \cite{Nord}:
\begin{equation}
<m^a_e>_t = \biggl<\frac{m_e c^2}{(1 -v^2/c^2)^{1/2}} -
\frac{e^2}{r} \biggl>_t/c^2 + \biggl<\frac{m_e v^2}{(1
-v^2/c^2)^{1/2}}-\frac{e^2}{r}\biggl>_t/c^2 = m_e + E/c^2 .
\end{equation}
We point out that, in Eq.(79), the averaged over time virial term
is zero due to the classical virial theorem. It is easy to show
that for non-relativistic case our Eqs.(78),(79) can be simplified
to
\begin{equation}
m^a_e = m_e + \biggl(\frac{m_e v^2}{2} - \frac{e^2}{r} \biggl)/c^2
+ \biggl(2 \frac{m_e v^2}{2}-\frac{e^2}{r}\biggl)/c^2
\end{equation}
and
\begin{equation}
<m^a_e>_t = m_e + \biggl<\frac{m_e v^2}{2} - \frac{e^2}{r}
\biggl>_t/c^2 + \biggl<2 \frac{m_e
v^2}{2}-\frac{e^2}{r}\biggl>_t/c^2 = m_e + E/c^2.
\end{equation}

\subsection{6.2. Active gravitational mass in quantum physics}

In this Subsection, we consider the so-called semiclassical theory
of gravity [14], where, in the Einstein's field equation,
gravitational field is not quantized but the matter is quantized:
\begin{equation}
R_{\mu \nu} - \frac{1}{2}R g_{\mu \nu} = \frac{8 \pi G}{c^4}
\bigl<\hat T_{\mu \nu} \bigl> .
\end{equation}
Here, $<\hat T_{\mu \nu}>$ is the expectation value of quantum
operator, corresponding to the stress-energy tensor. To make use
of Eq.(82), we have to rewrite Eq.(80) for electron active
gravitational mass using momentum, instead of velocity. Then, we
can quantize the obtained result:
\begin{equation}
\hat m^a_e = m_e +\biggl(\frac{{\bf \hat
p}^2}{2m_e}-\frac{e^2}{r}\biggl)/c^2 + \biggl(2\frac{{\bf \hat
p}^2}{2m_e}-\frac{e^2}{r}\biggl)/c^2.
\end{equation}
Note that Eq.(83) represents electron active gravitational mass
operator. As directly follows from it, the expectation value of
electron active gravitational mass can be written as
\begin{equation}
<\hat m^a_e> = m_e +\biggl< \frac{{\bf \hat
p}^2}{2m_e}-\frac{e^2}{r}\biggl>/c^2 + \biggl<2\frac{{\bf \hat
p}^2}{2m_e}-\frac{e^2}{r}\biggl>/c^2,
\end{equation}
where third term is the virial one.

\subsubsection{6.2.1. Equivalence of the expectation values of active
gravitational mass and energy for stationary quantum states}

Now, we consider a macroscopic ensemble of hydrogen atoms with
each of them being in the n-th energy level. For such ensemble,
the expectation value of the mass (83) is
\begin{equation}
<\hat m^a_e> = m_e + \frac{E_n}{c^2}.
\end{equation}
In Eqs.(84),(85), we take into account that the expectation value
of the virial term is equal to zero in stationary quantum states
due to the quantum virial theorem \cite{Park}. Thus, we can make
the following important conclusion: in stationary quantum states,
active gravitational mass of a composite quantum body is
equivalent to its energy at a macroscopic level
\cite{Lebed-1,Lebed-2}.

\subsubsection{6.2.2. Inequivalence between active gravitational mass and
energy for macroscopic coherent ensemble of quantum superpositions
of stationary states}

Below, we introduce the simplest macroscopic coherent ensemble of
quantum superpositions of the following stationary states in a
hydrogen atom,
\begin{eqnarray}
\Psi (r,t) = \frac{1}{\sqrt{2}} \exp \biggl( -i \frac{m_e c^2
t}{\hbar} \biggl) \biggl[&& \Psi_1(r) \exp\biggl( -i
\frac{E_1t}{\hbar} \biggl)
\nonumber\\
&&+ \exp(i \alpha) \Psi_2(r) \exp\biggl( -i\frac{E_2t}{\hbar}
\biggl) \biggl],
\end{eqnarray}
where $\Psi_1(r)$ and $\Psi_2(r)$ are the normalized wave
functions of the ground state (1S) and first excited state (2S),
respectively. We stress that it is possible to create the coherent
superposition, where $\alpha = const$ for all macroscopic
ensemble, by using lasers. It is easy to show that the
superposition (86) corresponds to the following constant
expectation value of energy in the absence of gravitational field,
\begin{equation}
<E> = m_e c^2 + \frac{E_1+E_2}{2}.
\end{equation}
Nevertheless, as seen from Eq.(84), the expectation value of
electron active gravitational mass operator for the wave function
(86) is not constant and exhibits time-dependent oscillations:
\begin{equation}
<\hat m^a_e> = m_e + \frac{E_1+E_2}{2 c^2} + \frac{V_{1,2}}{c^2}
\cos \biggl[ \alpha + \frac{ (E_1-E_2)t}{\hbar} \biggl],
\end{equation}
where $V_{1,2}$ is matrix element of the virial operator,
\begin{equation}
V_{1,2} = \int \Psi_1(r) \ \biggl(2\frac{{\bf \hat
p}^2}{2m_e}-\frac{e^2}{r}\biggl) \ \Psi_2(r) \ d^3{\bf r} \ ,
\end{equation}
between the above-mentioned two stationary quantum states. It is
important that the oscillations (88),(89) directly demonstrate
breakdown of the equivalence between the expectation values of
active gravitational mass and energy for coherent quantum
superpositions of stationary states \cite{Lebed-1,Lebed-2}. We pay
attention to the fact that such quantum time-dependent
oscillations are very general and are not restricted by the case
of a hydrogen atom. They are of a pure quantum origin and do not
have classical analogs.

\subsection{6.3. Experimental aspects}
In this short Subsection, we suggest an idealized experiment,
which allows to observe quantum time-dependent oscillations of the
expectation values of active gravitational mass (88). In
principle, it is possible to create a macroscopic ensemble of the
coherent quantum superpositions of electron stationary states in
some gas with high density. It is important that these
superpositions have to be characterized by the feature that each
atom (or molecule) has the same phase difference between two wave
function components, $ \Psi_1(r)$ and $ \Psi_2(r)$. In this case,
the macroscopic ensemble of the atoms (or molecules) generates
gravitational field, which oscillates in time similar to Eq.(88),
which, in principle, can be measured. It is important to use such
geometrical distributions of the molecules and a test body, where
oscillations (88) are "in phase" and, thus, do not cancel each
other.

\section{7. Summary}
In conclusion, in the review, we have discussed in detail breakdown of
 the equivalence between active and passive gravitational masses of
 an electron and its energy in a hydrogen atom. We stress that the
 considered phenomena are very general and are not restricted by atomic
 physics and the Earth's gravitational field. In other words, the above
 discussed phenomena exist for any quantum system with internal degrees of
 freedom and at any gravitational field.
 In this review, we also have improved several drawbacks of the original
 pioneering works.

\section*{Acknowledgments}

We are thankful to N.N. Bagmet (Lebed), V.A. Belinski, Steven
Carlip, Fulvio Melia, Jan Rafelski, Douglas Singleton, Elias Vagenas,
and V.E. Zakharov for fruitful and useful discussions.

\end{document}